\documentclass[twocolumn]{article}
\usepackage{amsmath}
\usepackage{amssymb}
\usepackage{graphicx}
\usepackage{hyperref}
\usepackage{listings}
\usepackage{xcolor}
\usepackage{tikz}
\usepackage{float}
\usepackage[utf8]{inputenc}

\usepackage{listings}
\usepackage{xcolor}

\definecolor{codegreen}{rgb}{0,0.6,0}
\definecolor{codeblue}{rgb}{0.0, 0.0, 0.6}
\definecolor{codegray}{rgb}{0.5,0.5,0.5}
\definecolor{codepurple}{rgb}{0.58,0,0.82}
\definecolor{backcolour}{rgb}{0.95,0.95,0.92}

\lstdefinelanguage{Rust}{
    morekeywords={abstract,alignof,as,become,box,break,const,continue,crate,do,else,enum,extern,false,final,fn,for,if,impl,in,let,loop,macro,match,mod,move,mut,offsetof,override,priv,proc,pub,pure,ref,return,Self,self,static,struct,super,trait,true,type,typeof,unsafe,unsized,use,virtual,where,while,yield,Vec},
    sensitive=true,
    morecomment=[l]{//},
    morecomment=[s]{/*}{*/},
    morestring=[b]',
    morestring=[b]''
}

\title{Polynomial Methods for Ensuring Data Integrity in Financial Systems}
\author{Ignacio Brasca}

\begin{document}
\maketitle

\section{Introduction}

In any sufficiently complex system, such as any vast financial system with any practical user base, ensuring data integrity across $K$ data points is crucial.

Furthermore, to satisfy correctness during usage, we should ensure that $k_n$ data points are verified and correct for consumption while $K$ is correct as a whole, without any discrepancies in expected values. 

Values falling out of valid ranges can cause distress among consumers of this data, leading to questions directed at platform designers, resulting in wasted time trying to recover original data points used, mitigating data loss, and validating inputs across $I$ indicators where $I$ is the number of total indicators used across the platform. (An indicator is a set of $k_n$ configured in a way that an operation produces a result consumed by the end-consumer.)

To address these issues, we propose implementing an algorithm that relies on the classical Lagrange Interpolation to help maintain correctness in our set of $I$ by using polynomials through known data points. 

\section{Background}

Indicators ($I_n$) play a fundamental role in any financial system where, for each specific $I_n$, a set $\{ k_1, k_2, \ldots, k_n \}$ could be in use. 

\begin{equation}
I_n = \sum_{i=0}^{k_n} f(k_i)
\end{equation}

where $f(k_i)$ is a function that takes a data point and produces a result. This is called an $operation$ and is used to calculate any measure required to obtain $I_n$.

An indicator, in conclusion, is a function that takes a set of data points and produces a result used in the calculation of other indicators or as an individual indicator ready for consumption. 

\subsection*{Polynomials}

The usage of polynomials to maintain consistency is well-discussed in the literature. Algorithms such as Reed-Solomon codes, used in error detection and correction, rely on polynomials to ensure data integrity \cite{cormen2009introduction}, \cite{macwilliams1977theory}, \cite{lidl1986introduction}.

This, in addition to the simplicity of polynomial representation, makes it a good candidate for ensuring data integrity.

Additionally, polynomials are key in solving complex problems in mathematics by breaking them down into simpler functions that can be recombined to form the original problem. Given any polynomial \( P(x) \), there exists a unique function \( y = \phi(t) \) that satisfies the differential equation:

\begin{equation}
y' + p(t)y = g(t), \quad y(t_0) = y
\end{equation}

This guarantees the existence and uniqueness of the polynomial for given points \cite{rana2002introduction}.

Knowing \textbf{existence} and \textbf{uniqueness} from a set of points, we can rely on that to understand how we can use this to ensure the integrity of our data points.

Shamir Secret Sharing \cite{shamir1979how} is another example of a cryptographic algorithm that relies on polynomial interpolation to recover a secret from a set of shares, leveraging the same properties of polynomials described later in this document.

\subsection*{Lagrange Interpolation}

Lagrange Interpolation \cite{burden2001numerical} is a method used to find a polynomial \( P(x) \) that passes through a set of given points \( (x_0, y_0), (x_1, y_1), \ldots, (x_n, y_n) \). The polynomial \( P(x) \) is constructed as a linear combination of basis polynomials:

\begin{equation}
P(x) = \sum_{i=0}^{n} y_i L_i(x)
\end{equation}

where \( L_i(x) \) are the Lagrange basis polynomials defined as:

\begin{equation}
L_i(x) = \prod_{\substack{0 \le j \le n \\ j \ne i}} \frac{x - x_j}{x_i - x_j}
\end{equation}

\subsubsection*{Lagrange Interpolation: An example}

Consider the points \( (1, 2) \), \( (2, 3) \), and \( (3, 5) \). Using Lagrange Interpolation, the basis polynomials are:

\begin{align}
L_0(x) &= \frac{(x - 2)(x - 3)}{(1 - 2)(1 - 3)} = \frac{(x - 2)(x - 3)}{2} \\
L_1(x) &= \frac{(x - 1)(x - 3)}{(2 - 1)(2 - 3)} = -(x - 1)(x - 3) \\
L_2(x) &= \frac{(x - 1)(x - 2)}{(3 - 1)(3 - 2)} = \frac{(x - 1)(x - 2)}{2}
\end{align}

The interpolating polynomial \( P(x) \) is given by:

\begin{equation}
P(x) = \sum_{i=0}^{2} y_i L_i(x)
\end{equation}

Simplifying, we obtain:

\begin{equation}
P(x) = \frac{1}{2}x^2 - \frac{1}{2}x + 2
\end{equation}

This polynomial passes through the points \( (1, 2) \), \( (2, 3) \), and \( (3, 5) \).

\section{Application}

The application of polynomials in our context allows us to recover and validate data points, ensuring continuity and uniqueness of the results across all inputs.

We describe a ${k_1, k_2, \ldots, k_n}$ set of data points used to generate a polynomial from a list of arguments used in the $I_n$ indicator.

Suppose we have a set of data points $(x_0, y_0), (x_1, y_1), \ldots, (x_n, y_n)$ to compute an indicator $I_n$.

Furthermore, we take a subset of these original data points and assume \([x_0 = 0, x_1 = 1, x_2 = 2, \ldots, x_n = n]\), where the value of \(f(x)\) at each \(x_i\) corresponds to the original data point's value. For example, \(f(0) = X\), \(f(1) = Y\), \(f(2) = Z\), and so on up to \(f(n) = n\).

From there, we can start describing a $f(x)$ polynomial from a set of points first defined by our data explanation from $k_n$ sample data points. After presenting a set of points, already defined, we can start interpolating the polynomial from the set of those same exact points.

Giving us a list of $(x_0, y_0), (x_1, y_1), \ldots, (x_n, y_n)$, defined across $(0, f(0)), (1, f(1)), \ldots, (n, f(n))$ from which we now have a polynomial that describes the function $f(x)$:

\begin{equation}
    f(x) = \sum_{i=0}^{n} y_i L_i(x)
\end{equation}

where $L_i(x)$ is the Lagrange basis polynomial defined as:

\begin{equation}
L_i(x) = \prod_{\substack{0 \le j \le n \\ j \ne i}} \frac{x - x_j}{x_i - x_j}
\end{equation}

Using $f(x)$, we can now generate a set of $m$ parity blocks that can be used to recover the original data points in case of data loss or corruption (as long as we store those blocks in a different place, see section \ref{sec:storing_data_points}).

Ideally, one can generate as many $k$ (where $k$ is the number of points used in the first place) parity blocks as needed; however, we can generate as many $m$ blocks as we want and store them across different data storages.

One caveat is that we can reconstruct the original value as long as we have at least $k$ points, where $k$ is the threshold number of points required to reconstruct the polynomial, and $n$ is $deg(P(x))$, which is $n = k - 1$.

This ensures original $k$ data points can be reconstructed from the parity blocks, maintaining data integrity and consistency without the need to store the original data points in multiple locations.

\subsection*{Example: Carbon footprint calculation}

One notable application is the calculation of the carbon footprint from the total investments of a portfolio~\cite{esma2019technical}:

\begin{equation}
\text{carbon footprint} = \frac{\text{total scope emissions}}{\text{total value of investments}}
\end{equation}

Breaking this down, the total scope emissions can be derived from individual companies' emissions. By using Lagrange Interpolation, we can ensure that the calculated carbon footprint is consistent and accurate.

For three companies with emissions:
\begin{align}
\text{Company A:} & \quad 300 \text{ tonnes} \\
\text{Company B:} & \quad 400 \text{ tonnes} \\
\text{Company C:} & \quad 300 \text{ tonnes} \\ 
\text{Total Value:} & \quad 3000 \text{ EUR}
\end{align}

We can define the polynomial \( P(x) \) denoted by the set of points: \( (1, 300) \), \( (2, 400) \), \( (3, 300) \), \( (4, 3000) \). The Lagrange Interpolation polynomial is:

\begin{equation}
P(x) = \frac{1}{6}x^3 - \frac{3}{2}x^2 + \frac{11}{6}x - 100
\end{equation}

By interpolating these values, we can recover the original carbon footprint even if some data points are lost. 

\section{Concerns and Data Recovery}

\subsection{Concerns}

One primary concern is the possibility of data corruption or loss. To mitigate this, we store data points in multiple locations and use redundancy to recover lost data. The method also ensures that any interpolation is reversible, allowing us to verify the integrity of the data. This redundancy is analogous to techniques used in Reed-Solomon error correction codes, which can correct multiple errors in data blocks~\cite{peterson1972error},~\cite{lin1983error}.

\subsection{Data Recovery Example}

To demonstrate data recovery, consider the earlier example of the carbon footprint calculation. If we lose some data points, we can use the remaining points and the interpolating polynomial to reconstruct the missing values. This is done using the Lagrange Interpolation formula and the stored data points.

\section{Storing Data Points}\label{sec:storing_data_points}

To further enhance data integrity, we store data points in a secondary data storage. This ensures that even in the case of primary storage failure, the data can be recovered from the secondary storage. The steps include:

\begin{enumerate}
    \item Prepare data parts for \( k \) data points.
    \item Construct the polynomial using Lagrange Interpolation.
    \item Sample additional points (parity blocks) and store them in secondary storage.
    \item Use the stored data and parity blocks to recover the original values if needed.
\end{enumerate}

This approach is similar to RAID 6 storage systems, which use Reed-Solomon codes to provide fault tolerance and data recovery capabilities~\cite{patterson1988case},~\cite{sudan1997decoding}.

\subsection{Example}

\begin{lstlisting}[language=Rust, caption={Pseudo code for storing data points}, label={lst:store_data_points}]
fn store_data_points(data_points: Vec<f64>) {
    // construct polynomial from data points
    let polynomial = polynomial(&data_points);

    // sample parity blocks
    let parity_blocks = sample_parity(&polynomial);

    // store data points and parity blocks
    store_in_primary_storage(&data_points);
    store_in_secondary_storage(&parity_blocks);
}

fn recover_data_points(data_points: Vec<f64>) -> Vec<f64> {
    // check if data points exist, if not, recover from backup
    let data_points = if !in_primary_storage() {
        retrieve_from_backup()
    } else {
        retrieve_from_primary()
    };

    // construct polynomial from available data points
    let polynomial = construct(&data_points);

    polynomial.interpolate(&data_points)
}
\end{lstlisting}

\section{Use Cases in Other Fields}

\subsection{Medical Imaging}

In medical imaging, accurate data reconstruction is crucial. Lagrange Interpolation can be used to fill in missing or corrupted pixel values in MRI and CT scans, ensuring accurate and reliable images for diagnosis~\cite{mceliece1977theory}.

\subsection{Climate Modeling}

Climate models rely on vast amounts of data from various sources. Lagrange Interpolation can help in interpolating missing data points from temperature, humidity, and other climatic variables, ensuring the models are accurate and robust~\cite{cormen2009introduction}.

\subsection{Engineering Design}

In engineering design, particularly in finite element analysis, interpolating values at various points in a mesh is essential. Lagrange Interpolation can provide accurate approximations of physical properties across the mesh~\cite{macwilliams1977theory}.

\section{Secondary Database for Data Recovery}

Consider a scenario where we have a critical equation used in financial forecasting:

\begin{equation}
F(t) = a \cdot e^{bt} + c \cdot \sin(dt)
\end{equation}

The coefficients \(a, b, c, d\) are stored in a primary database. In the event of a database failure, we can recover these coefficients using Lagrange Interpolation from a secondary database where interpolated parity blocks are stored. If the primary database fails or is corrupted, the following steps can be used to recover the coefficients:

\subsection{Recovery Procedure}

\textbf{Step 1: Check Primary Database Accessibility}

First, determine whether the primary database is accessible. This involves checking the network connectivity, server status, and database health. Ensuring the primary database is online and functioning correctly is crucial before proceeding to data retrieval.

\textbf{Step 2: Retrieve Parity Blocks from Secondary Database}

If the primary database is not accessible, initiate the retrieval process for parity blocks stored in the secondary database. These parity blocks are essential for reconstructing the missing data and are typically stored in a fault-tolerant manner.

\textbf{Step 3: Reconstruct Missing Coefficients Using Lagrange Interpolation}

Utilize the retrieved parity blocks and the known data points to perform Lagrange Interpolation. This mathematical technique will help in reconstructing the missing coefficients, ensuring that the recovered data is accurate and reliable.

\textbf{Step 4: Validate Recovered Coefficients}

Finally, validate the reconstructed coefficients by comparing them against known data points. This step is crucial for verifying the accuracy and integrity of the recovered data, ensuring consistency across the system.

\section{Conclusion}

The technique described here can be used to ensure data integrity across a set of $I$ indicators and maintain consistency in a system where traceability of the $I_n$ in use is critical. 

This technique can be applied in any critical system where robustness is mandatory. By relying on the mathematical properties of the Lagrange Interpolation, we can be sure our function will be well-defined in the domain of $\mathbb{R}$. 

Storing parity blocks in secondary databases allows us to ensure from them $k_{parity}$ and discard corrupted (or untrusted) data points. These parity blocks enable us to trace back across the domain and recover the set $\{(0, f(0)), (1, f(1)), \ldots, (n, f(n))\}$ data points used to generate the polynomial $f(x)$ in the first place. This remains true even if all the original data points are lost or corrupted, thanks to extrapolation and reliance on the uniqueness and existence of the polynomial $f(x)$ in the domain of $\mathbb{R}$ \cite{rudin1976principles}.

This approach provides fault tolerance (up to $m$ parity blocks) and data recovery capabilities. By implementing this technique, we can enhance data integrity and ensure the continuity of critical operations in various domains.

\section{Future Work}
Future work includes the implementation of the described technique in real-world scenarios and the evaluation of its performance in terms of data recovery, accuracy, and computational efficiency. 

\section{References}

\bibliographystyle{acm}
\bibliography{references}

\end{document}